\documentclass[aps,prl,superscriptaddress,floatfix,showpacs]{revtex4}
\usepackage{graphicx}
\usepackage{bm}
\usepackage{epsfig,subfigure}

\begin{document}
\newcounter{TC}
\newcommand{\useTC}[1]{\refstepcounter{TC}\label{#1}\arabic{TC}}
\newcommand{\MeVcsq}{${\rm MeV}/c^2$}
\newcommand{\GeVc}{${\rm GeV}/c$}
\newcommand{\EGEV}{${\rm GeV}/c$}
\newcommand{\PGEV}{${\rm GeV}^2/c^2$}
\newcommand{\MGEV}{${\rm GeV}/c^2$}


\title{Confirmation of  the $ 1^{-+}$  Meson Exotics in the  $ \eta \pi^0$ System }

\author{G.~S.~\surname{Adams}}
\affiliation{Department of Physics, Rensselaer Polytechnic Institute, Troy, New York 12180}
\author{T.~\surname{Adams}}
\altaffiliation[Present address: ]{Department of Physics, Florida State University, Tallahassee, FL 32306}
\affiliation{Department of Physics, University of Notre Dame, Notre Dame, Indiana 46556}
\author{Z.~\surname{Bar-Yam}}
\affiliation{Department of Physics, University of Massachusetts Dartmouth, North Dartmouth, Massachusetts 02747}
\author{J.~M.~\surname{Bishop}}
\affiliation{Department of Physics, University of Notre Dame, Notre Dame, Indiana 46556}
\author{V.~A.~\surname{Bodyagin}}
\altaffiliation{Deceased}
\affiliation{Nuclear Physics Institute, Moscow State University, Moscow, Russian Federation 119899}
\author{D.~S.~\surname{Brown}}
\altaffiliation[Present address: ]{Department of Physics, University of Maryland, College Park, MD 20742}
\affiliation{Department of Physics, Northwestern University, Evanston, Illinois 60208}
\author{N.~M.~\surname{Cason}}
\affiliation{Department of Physics, University of Notre Dame, Notre Dame, Indiana 46556}
\author{S.~U.~\surname{Chung}}
\affiliation{Physics Department, Brookhaven National Laboratory, Upton, New York 11973}
\author{J.~P.~\surname{Cummings}}
\affiliation{Department of Physics, Rensselaer Polytechnic Institute, Troy, New York 12180}
\author{A.~I.~\surname{Demianov}}
\affiliation{Nuclear Physics Institute, Moscow State University, Moscow, Russian Federation 119899}
\author{K.~\surname{Danyo}}
\affiliation{Physics Department, Brookhaven National Laboratory, Upton, New York 11973}
\author{J.~P.~\surname{Dowd}}
\affiliation{Department of Physics, University of Massachusetts Dartmouth, North Dartmouth, Massachusetts 02747}
\author{P.~\surname{Eugenio}}
\affiliation{Department of Physics, Florida State University, Tallahassee, FL 32306}
\author{X.~L.~\surname{Fan}}
\affiliation{Department of Physics, Northwestern University, Evanston, Illinois 60208}
\author{A.~M.~\surname{Gribushin}}
\affiliation{Nuclear Physics Institute, Moscow State University, Moscow, Russian Federation 119899}
\author{R.~W.~\surname{Hackenburg}}
\affiliation{Physics Department, Brookhaven National Laboratory, Upton, New York 11973}
\author{M.~\surname{Hayek}}
\altaffiliation[Permanent address: ]{Rafael, Haifa, Israel}
\affiliation{Department of Physics, University of Massachusetts Dartmouth, North Dartmouth, Massachusetts 02747}
\author{J.~\surname{Hu}}
\altaffiliation[Present address: ]{TRIUMF, Vancouver, B.C., V6T 2A3, Canada}
\affiliation{Department of Physics, Rensselaer Polytechnic Institute, Troy, New York 12180}
\author{E.~I.~\surname{Ivanov}}
\affiliation{Department of Physics, Idaho State University, Pocatello, ID 83209}
\author{D.~\surname{Joffe}}
\affiliation{Department of Physics, Northwestern University, Evanston, Illinois 60208}
\author{W.~\surname{Kern}}
\affiliation{Department of Physics, University of Massachusetts Dartmouth, North Dartmouth, Massachusetts 02747}
\author{E.~\surname{King}}
\affiliation{Department of Physics, University of Massachusetts Dartmouth, North Dartmouth, Massachusetts 02747}
\author{O.~L.~\surname{Kodolova}}
\affiliation{Nuclear Physics Institute, Moscow State University, Moscow, Russian Federation 119899}
\author{V.~L.~\surname{Korotkikh}}
\affiliation{Nuclear Physics Institute, Moscow State University, Moscow, Russian Federation 119899}
\author{M.~A.~\surname{Kostin}}
\affiliation{Nuclear Physics Institute, Moscow State University, Moscow, Russian Federation 119899}
\author{J.~\surname{Kuhn}}
\affiliation{Department of Physics, Rensselaer Polytechnic Institute, Troy, New York 12180}
\author{J.~M.~\surname{LoSecco}}
\affiliation{Department of Physics, University of Notre Dame, Notre Dame, Indiana 46556}
\author{M.~\surname{Lu}}
\affiliation{Department of Physics, Rensselaer Polytechnic Institute, Troy, New York 12180}
\author{L.~V.~\surname{Malinina}}
\affiliation{Nuclear Physics Institute, Moscow State University, Moscow, Russian Federation 119899}
\author{J.~J.~\surname{Manak}}
\affiliation{Department of Physics, University of Notre Dame, Notre Dame, Indiana 46556}
\author{M.~\surname{Nozar}}
\altaffiliation[Present address: ]{Thomas Jefferson National Accelerator Facility, Newport News, Virginia 23606}
\affiliation{Department of Physics, Rensselaer Polytechnic Institute, Troy, New York 12180}
\author{C.~\surname{Olchanski}}
\altaffiliation[Present address: ]{TRIUMF, Vancouver, B.C., V6T 2A3, Canada}
\affiliation{Physics Department, Brookhaven National Laboratory, Upton, New York 11973}
\author{A.~I.~\surname{Ostrovidov}}
\affiliation{Department of Physics, Florida State University, Tallahassee, FL 32306}
\author{T.~K.~\surname{Pedlar}}
\altaffiliation[Present address: ]{Laboratory for Nuclear Studies, Cornell University, Ithaca, NY 14853}
\affiliation{Department of Physics, Northwestern University, Evanston, Illinois 60208}
\author{L.~I.~\surname{Sarycheva}}
\affiliation{Nuclear Physics Institute, Moscow State University, Moscow, Russian Federation 119899}
\author{K.~K.~\surname{Seth}}
\affiliation{Department of Physics, Northwestern University, Evanston, Illinois 60208}
\author{N.~\surname{Shenhav}}
\altaffiliation[Permanent address: ]{Rafael, Haifa, Israel}
\affiliation{Department of Physics, University of Massachusetts Dartmouth, North Dartmouth, Massachusetts 02747}
\author{X.~\surname{Shen}}
\altaffiliation[Permanent address: ]{Institute of High Energy Physics, Bejing, China}
\affiliation{Department of Physics, Northwestern University, Evanston, Illinois 60208}
\affiliation{Thomas Jefferson National Accelerator Facility, Newport News, Virginia 23606}
\author{W.~D.~\surname{Shephard}}
\affiliation{Department of Physics, University of Notre Dame, Notre Dame, Indiana 46556}
\author{N.~B.~\surname{Sinev}}
\affiliation{Nuclear Physics Institute, Moscow State University, Moscow, Russian Federation 119899}
\author{D.~L.~\surname{Stienike}}
\affiliation{Department of Physics, University of Notre Dame, Notre Dame, Indiana 46556}
\author{J.~S.~\surname{Suh}}
\altaffiliation[Present address: ]{Department of Physics, Kyungpook National University, Daegu, Korea}
\affiliation{Physics Department, Brookhaven National Laboratory, Upton, New York 11973}
\author{S.~A.~\surname{Taegar}}
\affiliation{Department of Physics, University of Notre Dame, Notre Dame, Indiana 46556}
\author{A.~\surname{Tomaradze}}
\affiliation{Department of Physics, Northwestern University, Evanston, Illinois 60208}
\author{I.~N.~\surname{Vardanyan}}
\affiliation{Nuclear Physics Institute, Moscow State University, Moscow, Russian Federation 119899}
\author{D.~P.~\surname{Weygand}}
\affiliation{Thomas Jefferson National Accelerator Facility, Newport News, Virginia 23606}
\author{D.~B.~\surname{White}}
\affiliation{Department of Physics, Rensselaer Polytechnic Institute, Troy, New York 12180}
\author{H.~J.~\surname{Willutzki}}
\altaffiliation{Deceased}
\affiliation{Physics Department, Brookhaven National Laboratory, Upton, New York 11973}
\author{M.~\surname{Witkowski}}
\affiliation{Department of Physics, Rensselaer Polytechnic Institute, Troy, New York 12180}
\author{A.~A.~\surname{Yershov}}
\affiliation{Nuclear Physics Institute, Moscow State University, Moscow, Russian Federation 119899}


\collaboration{The E852 collaboration}
\noaffiliation


\date{\today}

\begin{abstract}
The exclusive reaction $\pi^- p \to \eta \pi^0 n$,  $\eta \to
\pi^+ \pi^- \pi^0$ at 18~GeV$/c$ has been studied
with a partial wave analysis on a sample of 23~492 $\eta \pi^0 n$ events from
BNL experiment E852.
A mass-dependent fit is consistent with a resonant hypothesis for the
$P_+$ wave, thus providing evidence for a neutral exotic meson with
$J^{PC} = 1^{-+}$, a mass of $1257  \pm 20  \pm
25$~MeV$/c^2$, and a width of $354 \pm 64 \pm 60$~MeV$/c^2$.
New interpretations of the meson exotics in neutral $\eta \pi^0$ system observed in E852 and Crystal Barrel experiments are discussed.

\end{abstract}


\maketitle


\section{Introduction.}
Exotic mesons with
$J^{PC}=0^{--},1^{-+},2^{+-},\ldots$ do not mix with quark-antiquark
mesons and thus offer a natural testing ground for QCD. Exotic
mesons have been discussed
\cite{jaffe,baclose,th:barnes,th:ip,qcdsum,th:bcs,th:clp,th:diquark,th:lgt,barnesmeson2000}
for many years but have only recently been observed experimentally.
The underlying structure of the negatively charged exotic state with
$J^{PC}=1^{-+}$ observed in this experiment
 \cite{thompson,long_paper} at 1400~MeV decaying into $\eta\pi^-$ is
not yet understood.

Two distinguishing characteristics of the $\eta \pi^0$ system make it an
excellent one to clarify the properties of this exotic state.
First, $C$-parity is a good quantum number in the $\eta \pi^0$ system,
unlike the $\eta \pi^-$ system.
Second, the production mechanism for the
charge exchange reaction $\pi^- p \to \eta \pi^0 n$ cannot involve
the exchange of an isospin $I = 0$ system such as the pomeron.

The Crystal Barrel  experiment  \cite {cbarrel_1} confirmed the
existence of resonant structure in the $\eta \pi^-$ system using
stopped antiprotons  in liquid deuterium in the reaction $\bar p n
\to \pi^- \pi^0 \eta$. Later this group analyzed  data on $\bar p p$
annihilation at rest into $\pi^0 \pi^0 \eta$  \cite {cbarrel_2} and
presented evidence  for an exotic $1^{-+}$ resonance in the $\eta
\pi^0$ system with $M=(1360\pm 25)$~MeV$/c^2$ and $\Gamma=(220\pm
90)$~MeV$/c^2$.

The $\eta \pi^0$ state  has been studied in the GAMS experiment
 \cite {Alde} with $\pi^- p \to \eta \pi^0 n$, $\eta
\to 2\gamma$, $\pi^0 \to 2\gamma$ at 32, 38 and 100~GeV$/c$.
The statistics of the 38~GeV$/c$ data was sufficient so that,
using the method of Sadovsky \cite{sadov},
they were able to present evidence for the exotic $\pi_1(1400)$.

The VES experiment also observed a peak in the $P_+$ wave of the $\eta \pi^0$ system
near  1400~MeV$/c^2$ \cite{ex:VESetapm}.
In  their most recent publication \cite{amelin},
using theoretical arguments the authors state that the peak can be
understood without requiring an exotic meson.

An analysis of E852 data with $\pi^- p \to \eta \pi^0
p$, $\eta \to 2\gamma$ was recently reported \cite{dzierba}.
A bump in the $P_+$ wave of the $\eta \pi^0$ system was observed at
$M(\eta \pi^0)=1272$~MeV$/c^2$ with a large width ($\Gamma =
660$~MeV$/c^2$) when fitting all the data using the method of
Ref.~\cite{long_paper}. For small $t'$, they observe a width of
$190$~MeV$/c^2$. The authors chose not to claim evidence for exotic
$\pi_1(1400)$ meson production.

The present analysis studies
$\pi^- p \to \eta \pi^0 n$, $\eta \to \pi^+ \pi^- \pi^0$
at 18~GeV$/c$ in E852.
The advantage of this mode over the
all-neutral final state is that the production vertex is
defined by charged tracks.
This improves the mass resolution as well
as the ability to require that the interaction took place within the
target.

\section{Experimental setup and data selection}

The data for this analysis were obtained
  at the Alternating Gradient
Synchrotron (BNL USA) in 1995.  Using an  18~GeV$/c$ $\pi^-$ beam
interacting in a liquid hydrogen target, a total of 750 million
triggers were acquired of which 108 million were of a type designed
to enrich the exclusive final state  $\pi^- p \to \pi^+ \pi^-
4\gamma n$. A total of 6 million events of this type were fully
reconstructed. The data were kinematically fit \cite{squaw} to
select events consistent with the $\pi^-\pi^+\pi^0 \pi^0 n$
hypothesis (with a confidence level of at least 0.01\%) yielding
about 4 million events. Of those, 85~228 events passed a mass cut
enhancing $\eta$ mesons, $m(\pi^-\pi^+\pi^0) < 0.65$~GeV/$c^2$, and
74~549 passed a cut to remove events passing through a
low-efficiency region in the drift chambers.
A final kinematic fit selected 23~492 events for the partial wave
analysis (PWA), which were consistent with the $\eta\pi^0
n,~\eta\rightarrow\pi^+\pi^-\pi^0$ hypothesis at a minimum
confidence level of 1\%.

\begin{figure}
\centering
\mbox{\subfigure{\epsfig{figure=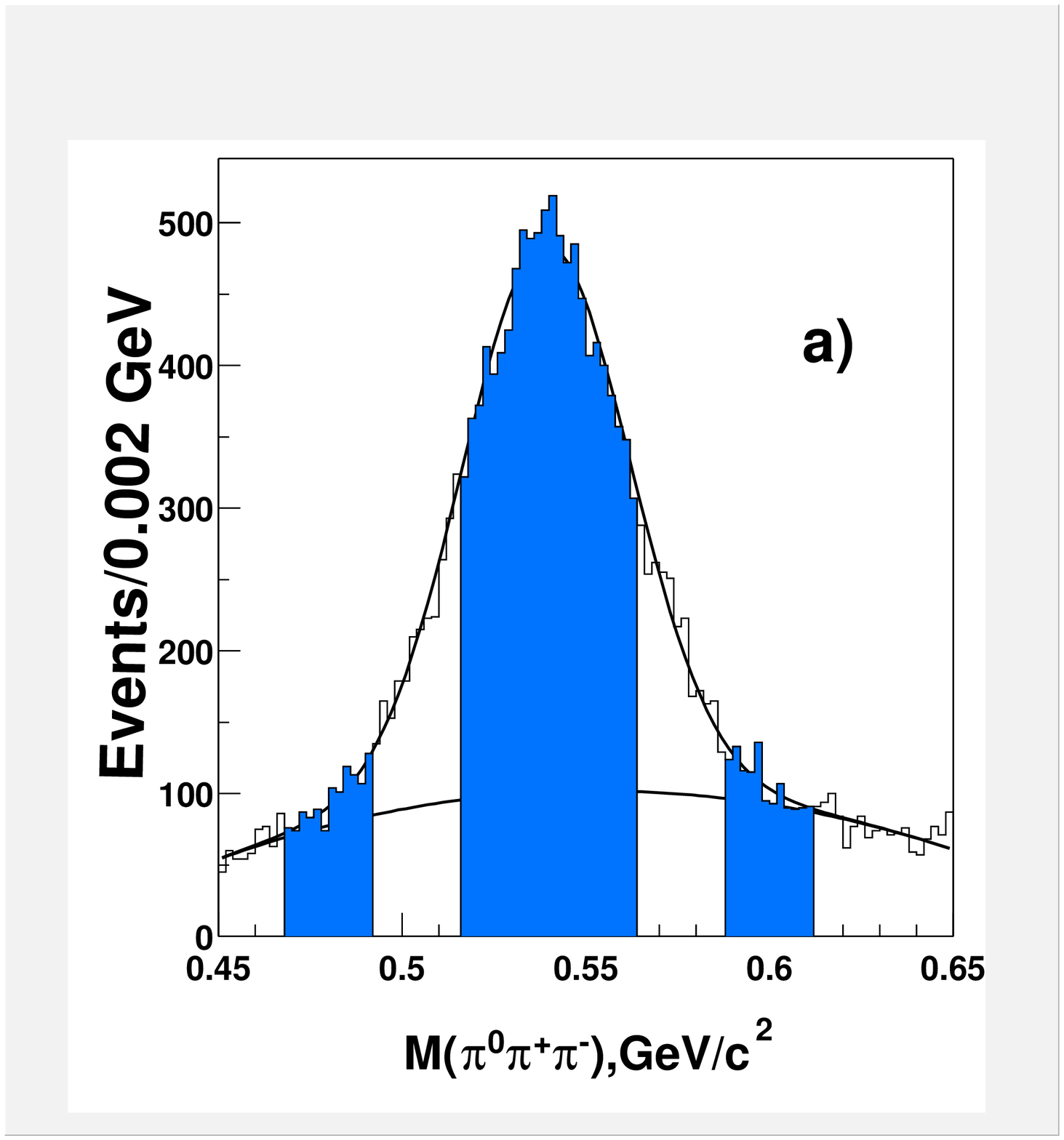,width=4.3cm}}
\subfigure{\epsfig{figure=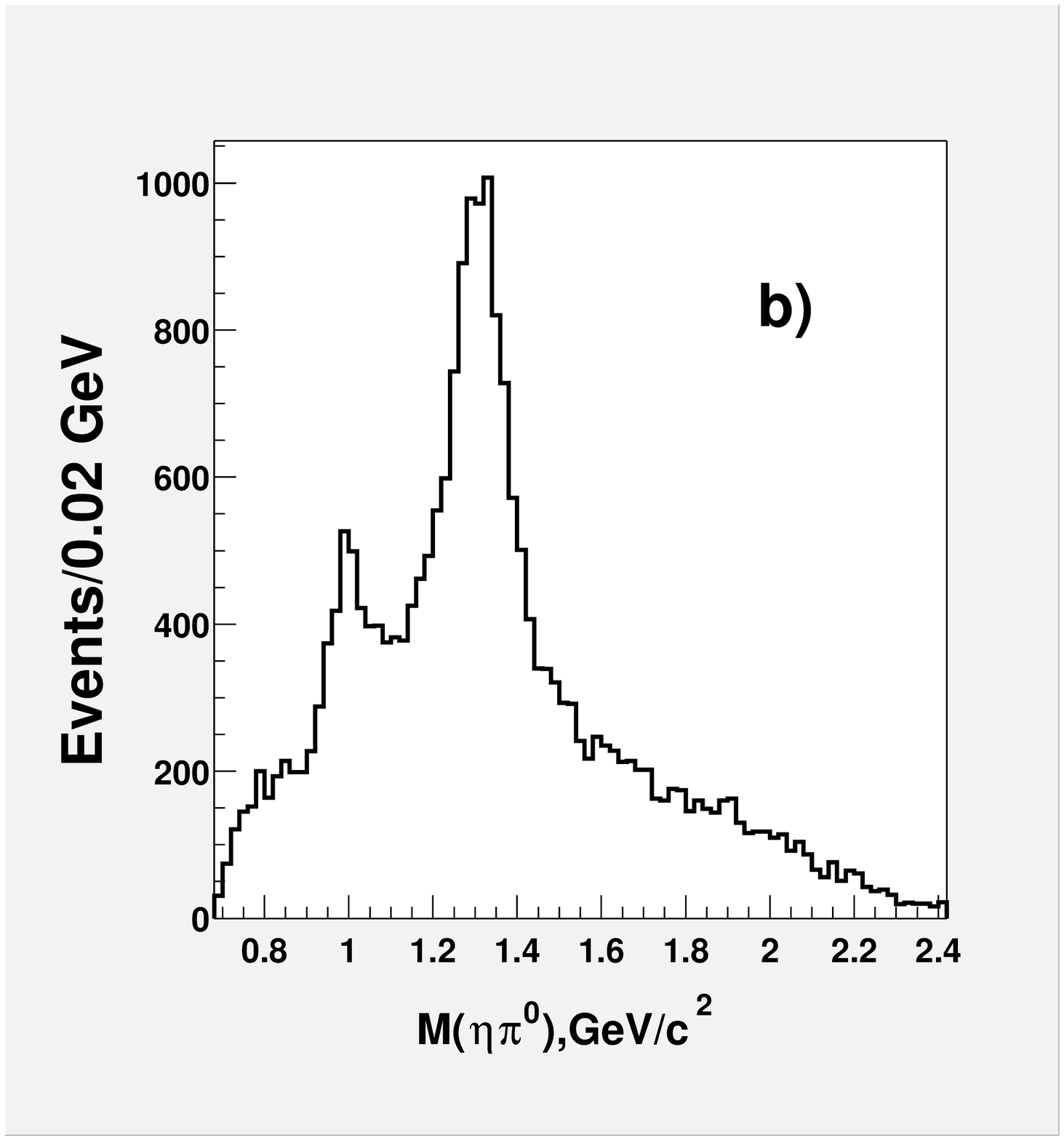,width=4.3cm}}}
\caption{\label{signal}(a) Fit of the $ \pi^+\pi^-\pi^0$ mass
distribution in the $\eta$ mass region.
There are two entries per event: one for each way to assign a
$\pi^0$ to the $\eta$ decay.
The $\eta$ signal region and the side-band regions are shown shaded. (b)
The uncorrected $\eta \pi^0$ effective-mass distribution for events
consistent with the  reaction $\pi^- p \to \eta \pi^0 n$}.
\end{figure}

A strong $\eta$ meson signal is observed in this final data sample
(Fig. \ref{signal}a) with a mass of $539.2 \pm 0.3$~MeV$/c^2$ and a
(resolution-dominated) width of $23.7 \pm 0.2$~MeV$/c^2$. The filled
regions in the figure indicate the signal region and the side-band
regions used in the analysis. In the $\eta$ signal region, the
signal-to-background ratio is about 4 to 1 for all $\eta \pi^0$
masses and 5 to 1 for $m(\eta \pi^0) > 1.1$~GeV/$c^2$. The $\eta
\pi^0$ mass spectrum shown in Fig.\ref{signal}b has two clear peaks:
the $a_0^0(980)$ and the $a_2^0(1320)$.

The non-$\eta$ background  was estimated  as a function of $\eta
\pi^0$ mass using the side-band and signal regions. The background
fraction varies between $24\%$ and $14\%$ going from lower to higher
mass in the region $0.78 < m(\eta \pi^0) < 1.74$~GeV/$c^2$.


The experimental acceptance was determined using a Monte Carlo event
sample generated with isotropic angular distributions in the
Gottfried-Jackson frame. The detector simulation was based on the
E852 detector simulation package SAGEN \cite{long_paper}. The
experimental acceptance was incorporated into the PWA by means of
Monte Carlo normalization integrals \cite{long_paper}. The
acceptance as a function of mass and as a function of $t'$ is flat.

\section{Partial Wave Analysis}

The partial-wave analysis (PWA) method described in
\cite{long_paper} (see also \cite{newbnl,th:SUtwo}) was used to study the
spin-parity structure of the $\eta \pi^{0}$ system in this data set.
The PWA was
carried out using the extended maximum likelihood method separately
in each mass bin in the mass region
between 0.78 and 1.74~GeV in mass bins of 0.04~GeV for $0 < |t'|
<1.0~($GeV$/c)^2$ using the likelihood function
\begin{equation}
   {\rm ln}{\cal L}\propto \sum^n_i\ {\rm ln} I(\Omega_i)
              -\int{\rm d}\Omega { \eta(\Omega) \,I(\Omega)}.
\label{f_pwa}
\end{equation}
Here $I(\Omega)$ is the predicted angular distribution,
$\eta(\Omega)$ is the angular acceptance, and the sum is over the
event sample.

The partial waves are parameterized by a set of five numbers:
$J^{PC} m^{\epsilon}$, where $J$ is the angular momentum, $P$ and
$C$ are the parity and the C-parity of the $\eta \pi^{0}$ system,
$m$ is the absolute value of the angular momentum projection and
$\epsilon$ is the reflectivity.
We use a simplified notation
where each partial wave is denoted by a letter indicating the $\eta
\pi^{0}$ system's angular momentum in standard spectroscopic
notation, and a subscript which can take the values $0,\;+$, or $-$,
for $m^{\epsilon}=0^{-}$, $1^{+}$, or $1^{-}$ respectively. We
assume that the contribution from partial waves with $m> 1$ is small
and can be neglected \cite{long_paper}.

The amplitudes used are the unnatural parity-exchange waves (UNPW)
$S_0$, $P_0$, $P_-$, $D_0$, $D_-$, and the natural parity-exchange
waves (NPW) $P_+$, $D_+$. The NPW waves interfere between themselves
as do the UNPW waves but the NPW waves do not interfere with the
UNPW waves. The $P_+$ wave would be an exotic $J^{PC}=1^{-+}$
(denoted by $\pi_1^0$) if the wave is resonant.

For each partial wave the complex production amplitudes were
determined from an extended maximum likelihood fit \cite{th:SUtwo}.
The spin 1/2 nature of the target proton leads to spin-flip and
spin-non-flip amplitudes and thus to a production spin-density
matrix with maximal rank two. The PWA fit presented in this paper
was carried out with the assumption that a spin-density matrix of
rank one was sufficient \cite{long_paper}. An isotropic incoherent
background was included. The magnitude of the background was fixed
as determined from the side bands.
We investigate the quality of the fits by comparing the moments of
the decay angular distributions $H(LM), L\leq 4$
\cite{long_paper,th:SUtwo}, of the data with those predicted by
Monte Carlo events generated with the fit amplitudes. We also
directly compare  the angular distributions for cos$(\theta_{GJ})$
and $\varphi_{TJ}$ between the data and those Monte Carlo events.
The quality of the fits is good.

  Since natural-parity exchange
(NPE) and unnatural-parity exchange (UNPE) amplitudes have different
$|t'|$ dependences, a fit to the $|t'|$ distribution using a
function of the form $N(t') = n_1 |t'| e^{-b_1|t'|} + n_2
e^{-b_2|t'|}$ was carried out to determine the relative
contributions of the two exchanges. The fitted parameters are
$b_1=(7.41\pm 0.08)(\mbox{GeV/c})^2,~b_2=(2.68\pm
0.07)(\mbox{GeV/c})^2,$ and thus the ratio of UNPE and NPE
contributions is equal to $ 0.71 \pm 0.03$. A value of about 70\%
for the ratio of UNPE to NPE at 18 GeV/c is expected based on
interpolation of experimental $\pi^- p$ charge exchange data between
3 and 40 $\mbox{GeV/c}$ in the Regge model \cite{sadov}.

\begin{figure}
\centering
\includegraphics[width=8.6cm]{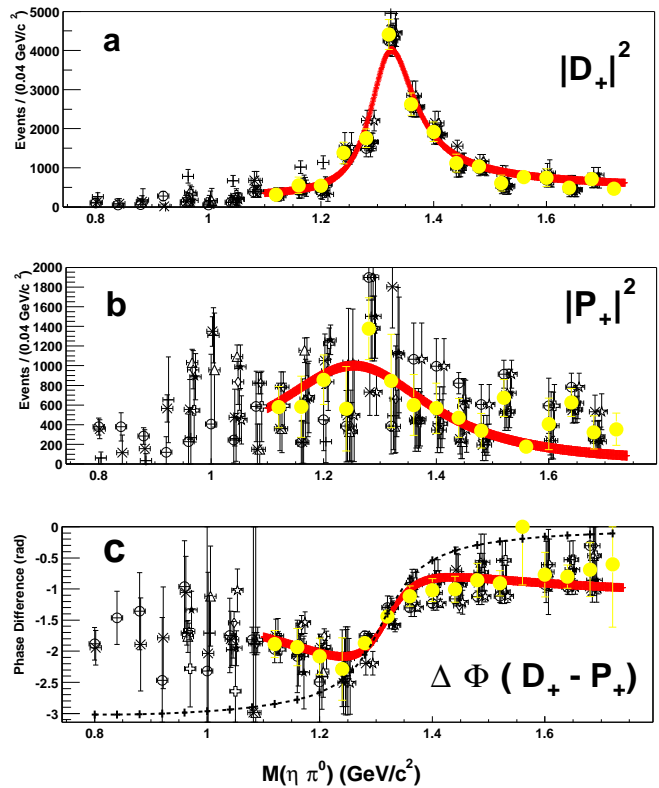}
\caption{\label{averag_NPW} The Partial Wave Analysis (PWA) and
Mass-Dependent Fit (MDF) results. The points shown in each mass bin
are the eight ambiguous PWA solutions. a) the $D_+$ wave intensity;
b) the $P_+$ wave intensity; and c) the relative phase between the
$P_+$ and $D_+$ waves. The  lines show the MDF results (method 2).
The average
PWA solution in each mass bin is plotted using grey (yellow) points. The
dotted line in (c) is the phase difference if the $P_+$ phase is
constant. }

\end{figure}

There is mathematical ambiguity in the description of a system of
two pseudo-scalar mesons~\cite{sadovsky_amb}. For our set of
amplitudes there are eight ambiguous solutions, each of which leads
to identical angular distributions. These solutions were found
analytically starting from one solution by means of the Barrelet
zeros method~\cite{th:SUtwo}. The eight solutions in each mass bin
are shown in Fig. \ref{averag_NPW} as a point for every ambiguous
solution. (In some cases the solutions are too close together to be
visible as separate solutions. This is particularly true in the mass
bins at $M=1.56$~{GeV/c}$^2$ and at $M=1.72$~{GeV/c}$^2$.) The
spread between the various ambiguous solutions dominates the
systematic uncertainty in the resonance parameter determination (see
below).


\section{Mass dependent fits}

To study the resonant structure in the partial waves, we used three
different procedures.
The first method utilizes a Mass Dependent Fit (MDF) of the average
solutions in the NPW sector. The PWA results in each mass bin were
averaged between ambiguous solutions \cite{long_paper}. The mass
dependence of  the $P_+$ and $D_+$ intensities as well as their
relative phase difference were then fit by relativistic Breit-Wigner
(BW) functions (in both the $P_+$ and $D_+$ waves) with
mass-dependent widths and Blatt-Weisskopf barrier
factors\cite{long_paper}. The mass and width of the $a_2^0$ are well
known and were fixed using values\cite{book:pdb}:
$M=1320$~{MeV/c}$^2$, and $\Gamma=120$~{MeV/c}$^2$. The width of the
$a_2(1320)$ includes the experimental mass resolution. There are
three free parameters in the fit of the $D_+$ intensity, $|D_+|^2$:
one for the magnitude and two parameterizing the smooth background
for the $D_+$ wave, as was done in \cite{long_paper}.

In the MDF of the $|P_+|^2$ distribution
and the relative phase $\Delta \Phi(D_+ - P_+)$
 there are four free parameters: three from the BW
function and one for the production phase (assumed constant).
 The fit was carried out in the mass interval
$1.1-1.74$~GeV$/c^2$. The resonant hypothesis for $D_+$ and $P_+$
waves with a mass-independent production phase gives a
$\chi^2/$DoF=1.14 for 28 degrees of freedom. The non-resonant
hypothesis (no phase variation for the $P_+$ wave) gives
$\chi^2/$DoF=3.02. It is clear from Fig. \ref{averag_NPW}c that a
single resonant phase for the $a_2(1320)$ (dotted line) with a
constant (non-resonant) $P_+$ wave is not satisfactory.

The contrast between the $D_+ - P_+$ phase variation and the $D_+$
phase variation with mass (Fig. \ref{averag_NPW}c, solid line) shows
clearly that the observed phase variation is consistent with
interference between two resonant waves. It's worth pointing out
that the mass dependence of the $D_+ - P_+$ relative phase would be
nearly flat if each wave were resonant
 with very similar masses and widths.

The $P_+$ resonant parameters from the fit with the average
solutions and the average error matrix \cite{long_paper} are: $M =
1265 \pm 21$~MeV/c$^2$ and $\Gamma = 411 \pm 64$~MeV/c$^2$.


The second method was similar to the first except that instead of
fitting the average solutions, a large number $(\simeq 10^3)$ of
randomly chosen combinations of ambiguous solutions in each mass bin
were used as input to the mass-dependent fit. The obtained
distributions of the mass and width of the $P_+$ resonance for those
fits with acceptable values of $\chi^2$ (that is, those with
$\chi^2/DoF < 2$) were then fitted by a Gaussian.  The mean values
of these distributions are: $M = 1257 \pm 20 $~MeV/c$^2$ and $\Gamma
= 354 \pm 64~$MeV/c$^2$. The  curves shown in Fig. \ref{averag_NPW}
are drawn using these mean values.

The RMS values of these distributions are $\sigma_M = 25$~MeV/c$^2$
and $\sigma_\Gamma = 58~$MeV/c$^2$. The systematic errors are
obtained using these RMS values. Our analysis shows that this spread
due to the different ambiguous solutions dominates the systematic
error.



The third method used was to carry out a Mass-Dependent Partial Wave
Analysis (MDPWA) \cite{long_paper}. In this procedure, an extended
maximum likelihood function is generalized to include not only the
angular distribution, but also the $\eta\pi^0$ mass distribution for
each wave. This  analysis is free from the problem of ambiguous
solutions but it is necessary to parameterize the mass dependence of
every partial wave (including the UNPWs) and all relative phases. We
use the same parametrization for the $D_+$ and $P_+$ waves as in the
first two methods. The mass dependence of the UNPW waves were chosen
to be polynomials of second order with constant phases except for
the $S_0$ wave.  The $S_0$ wave was fitted with a  BW function using
the $a_0(980)$ resonance parameters. The MDPWA results for the $P_+$
wave are $M = 1256  \pm 10$~MeV/c$^2$ and
 $\Gamma = 319 \pm 34$~MeV/c$^2$, consistent with the results of
 methods 1 and 2.

In \cite{long_paper} it was shown that a pure $D_+$ wave can
artificially induce a $P_+$ wave  due to the effects of finite
acceptance and resolution. This ``leakage" leads to a $P_+$ wave
that mimics the $D_+$ intensity and phase. In our case, the $P_+$
intensity would therefore have an intensity with the shape of the
$a_2(1320)$ and a $D_+ - P_+$ phase difference which doesn't depend
on mass. These features allowed us to include in the MDPWA fit a
leakage term with these features.  We observed that
the leakage contribution to the $P_+$ wave from the $D_+$ wave is
negligible.


Evidence for a resonance interpretation for the $P_+$ wave is
primarily the behavior of the $D_+ - P_+$ relative phase
(Fig.~\ref{averag_NPW}c).  Since the $D_+$ phase variation is well
known because of the $a_2(1320)$ production, it is clear that the
$P_+$ phase cannot be constant (see dotted curve in
Fig.~\ref{averag_NPW}c) and it is well-described by a BW phase
variation.


The ratio of the $P_+$ and  $D_+$  intensities in the range  $1.24 <
M(\eta \pi^0) < 1.34$~GeV is equal to $|P_+|^2/|D_+|^2 = 0.43 \pm
0.10$. This ratio is larger than that for the $\eta \pi^-$ system,
as reported in Ref. \cite{long_paper}.

The mass of the neutral exotic $1^{-+}$ state, decaying into $\eta
\pi^0$, observed here ($M = 1257 \pm 20 \pm 25$~MeV/c$^2 $) is lower
than the mass observed in the Crystal Barrel experiment \cite
{cbarrel_2} ($M = 1360 \pm 25$~MeV/c$^2 $)  by about 100~MeV
although the results are barely consistent within errors.   The
width measured here ($\Gamma = 354 \pm 64 \pm 58$~MeV/c$^2$) is also
consistent with that from the Crystal Barrel measurement ($\Gamma =
220 \pm 90$~MeV/c$^2$).


It should also be noted that our result is similar to those obtained
from the low-$t'$ fits in Ref.~\cite{dzierba}.  They obtained $M =
1301 \pm 14$~MeV/c$^2 $ and $\Gamma = 190 \pm 32 $~MeV/c$^2$ in one
analysis, and $M = 1386 \pm 32$~MeV/c$^2 $ and $\Gamma = 363 \pm 81
$~MeV/c$^2$ in a fit to the experimental moments. No systematic
errors were given for either method.

The lower mass found in our analysis may be a consequence of
interference between the resonant state and background in the $\eta
\pi^0$ system, some of which may be from rescattering between the
$\eta$ and the $\pi^0$.
It is also possible that two or more  resonant $1^{-+}$ states may
be present in the $\eta \pi^0$ decay channel in the mass interval
mass between 1200 and 1400 ~MeV as might be expected if the exotic
state is a four quark state.

The mass for the $\pi^0_1$ is some 100 MeV below the negatively
charged $\pi^-_1$ measured by Crystal Barrel \cite {cbarrel_1} and
BNL-E852 \cite{thompson,long_paper}. Of course, this does not
constitute compelling evidence for more than one  exotic state in
this mass region.  However, if we {\em assume} that there are two
$\eta\pi$ states at 1280 and at 1380 MeV respectively, then one
expects the lower-mass state to have a substantial branching ratio
into $\rho\pi$ (the reaction $\pi^-+p\to\eta\pi^0+n$ is mediated by
$\rho$ exchange), whereas the higher-mass state would couple to the
$f_2(1270)\pi$ channel or perhaps to the  $Pomeron+\pi$ channel (the
reaction $\pi^-+p\to\eta\pi^-+p$ is produced in part by the exchange
of the Pomeron).


\section{CONCLUSIONS}

Mass dependent fits of the $D_+$ and $P_+$ amplitudes and their
relative phase using three different methods (described above) all
lead to the conclusion that the $P_+$ wave is well-described by a
resonance hypothesis and is inconsistent with having a constant
phase.  The resonance parameters for the  observed $\pi_1^0$ are
given by $M = 1257 \pm 20 \pm 25$~MeV/c$^2 $ and $\Gamma = 354 \pm
64 \pm 58$~MeV/c$^2$.  Here the first error is statistical and the
second is systematic.  (We have chosen to take the resonant
parameters and errors from method 2.) This result, together with the
previous results from Crystal Barrel \cite {cbarrel_2} and E852
~\cite{dzierba} provide strong indications for one or more
spin-exotic mesons near 1400 MeV/c$^2$ decaying to $\eta \pi^0$.



We would like to express our appreciation to the members of the
MPS group,
and to the staffs of the AGS, BNL, and  the  collaborating
institutions for their  efforts. This research was supported in
part by the National Science Foundation, the US Department of
Energy, and the Russian Ministry for Education and Science. The
Southeastern Universities Research Association
(SURA) operates the Thomas Jefferson National Accelerator Facility
for the United States Department of Energy under contract
DE-AC05-84ER40150.



\end{document}